# Quantum Hall Effect Measurement of Spin-Orbit Coupling Strengths in Ultraclean Bilayer Graphene/WSe$_2$ Heterostructures

*Dongying Wang[1], Shi Che[1], Guixin Cao[1], Rui Lyu[2], Kenji Watanabe[3], Takashi Taniguchi[3], Chun Ning Lau[1], Marc Bockrath[1]\**

[1] Department of Physics, The Ohio State University, Columbus, OH 43210, USA

[2] Department of Physics and Astronomy, University of California, Riverside, CA 92521, USA

[3] National Institute for Materials Science, Namiki Tsukuba Ibaraki 305-0044 Japan.

ABSTRACT

We study proximity-induced spin-orbit coupling (SOC) in bilayer graphene/few-layer WSe$_2$ heterostructure devices. Contact mode atomic force microscopy (AFM) cleaning yields ultra-clean interfaces and high-mobility devices. In a perpendicular magnetic field, we measure the quantum Hall effect to determine the Landau level structure in the presence of out-of-plane Ising and in-plane Rashba SOC. A distinct Landau level crossing pattern emerges when tuning the charge density and displacement field independently with dual gates, originating from a layer-selective SOC proximity effect. Analyzing the Landau level crossings and measured inter-Landau level energy gaps yields the proximity induced SOC energy scale. The Ising SOC is ≈ 2.2 meV, 100 times higher than the intrinsic SOC in graphene, while its sign is consistent with theories predicting a dependence of SOC on interlayer twist angle. The Rashba SOC is ~15 meV. Finally, we infer the magnetic field dependence of the inter-Landau level Coulomb interactions. These ultraclean bilayer graphene/WSe$_2$ heterostructures provide a high mobility system with the potential to realize novel topological electronic states and manipulate spins in nanostructures.

Efforts to control spin and electronic ground state topology in devices has driven an intensive study of spin-orbit coupling in materials.[1, 2] Graphene is a two-dimensional (2D) material with excellent electronic properties but small spin-orbit coupling (SOC),[3-6] motivating efforts to increase it.[7-21] For example, adatoms such as hydrogen or transition metals increase graphene's SOC,[7, 8] however these add disorder and lower charge mobility.[7] In another approach, coupling graphene to 2D materials with strong SOC such as transition metal dichalcogenides also increases graphene's SOC,[9-16, 18, 19, 21] while the interfacing of two crystalline systems adds less intrinsic disorder. Such studies have shown weak antilocalization signatures of spin-orbit coupling[9, 10, 13-16] as well as spin Hall signatures.[12] However, electrical transport studies in high mobility devices on the quantum Hall effect are limited, which can act as a precise probe of the SOC in graphene, and potentially realize new topological ground states.[19, 20, 22] Moreover, while recent progress has been made by a combination of capacitance and some transport measurements,[22] inter-Landau level energy gaps have not been measured, nor the Rashba SOC or magnetic field dependence of the inter-Landau level Coulomb interactions.

Here we report coupling bilayer graphene (BLG) via layer stacking to $WSe_2$, a 2D semiconductor with strong SOC, in hexagonal BN (hBN) encapsulated devices. We obtain high mobility devices (~110,000 $cm^2V^{-1}s^{-1}$) by squeezing contaminants away from device areas using contact AFM.[23] Our devices are dual gated, allowing independent tuning of the charge density $n$ and perpendicular displacement field $D$. At low temperatures under a perpendicular magnetic field $B$, we observe the quantum Hall effect (QHE) with all degeneracies lifted for $D = 0$, resulting from inversion symmetry breaking by the $WSe_2$ layer contacting the graphene.

Applying a nonzero $D$ yields a finite interlayer potential difference, enabling us to observe Landau level (LL) crossings[24-28] that differ in $D$ values from isolated BLG.

To understand this behavior, we compare our results to a single-particle theory that includes an out-of-plane Ising SOC $\lambda$ as well as an in-plane Rashba SOC $\lambda_R$.[20] For the nearly degenerate zero energy Landau level octet, most crossings deviate from the single-particle picture due to Coulomb interactions, except for the $\nu = \pm 3$ crossings.[20, 22, 29] From these we extract $\lambda \sim -2.2$ meV, where the sign indicates the direction of the effective out-of-plane magnetic field from the SOC in the bilayer graphene at the K point of the Brillouin zone (with the opposite direction at the K' point).[20] consistent with recent theoretical work predicting that $\lambda$ can be negative depending on the graphene/WSe$_2$ twist angle.[30] To gain further insight into the device behavior, we measure the inter-LL gaps using temperature-dependent transport measurements. These gaps are potentially affected by both the Landau level width as well as Coulomb interactions. We therefore study the rate of gap closure with displacement field, which is independent of these within a Hartree-Fock model. Measuring these rates for several gaps, we find the best fit to the measured values using the Rashba SOC as a fit parameter, yielding $\lambda_R \approx 15$ meV.

Using this $\lambda_R$ value, we then compare the measured gap evolution with $B$ for $\nu = 5$, 6, and 7 to the single particle theory. The $\nu = 5$ and 7 gaps' behavior is close to the theory predictions, while the $\nu = 6$ gap shows significant deviations. From these deviations we determine the Coulomb interactions representing the exchange splitting for opposite spins for states deriving from the K and K' points. We find that the interaction energy scales vary approximately linearly with $B$ over the measured range, consistent with previously found behavior.[31, 32] The $\nu = 5$ and $\nu = 7$ gaps differ slightly, which may originate from differences in exchange energy due to the Rashba coupling. Finally, we compare the crossing points to the single particle model, using $\lambda_R$ as a fit parameter, which yields 20 meV, somewhat larger but in reasonable agreement to that found using the gap closing rates. This suggests that the crossing points not too strongly affected by Coulomb interactions.

Our devices were fabricated using a dry transfer and stacking method.[33] Both BLG and few-layer WSe$_2$ flakes were first exfoliated from bulk crystals, and then stacked and encapsulated between atomically flat hBN layers. The interlayer BLG/WSe$_2$ twist angle was ~15° as determined by the layer edge alignment. After transfer, the stack was deposited on a Si wafer capped with 285 nm of SiO$_2$ and vacuum annealed at 360°C for one hour. To further promote the interlayer coupling of BLG and WSe$_2$, we used an AFM tip to controllably squeeze out trapped contaminants from the stack.[23] A clean region was identified by AFM and the rest was etched by a mixture of CHF$_3$ (40 sccm) and O$_2$ (4 sccm) gas into a Hall bar geometry. After the etch process, Cr/Au (5nm/70nm) electrodes were deposited to make one-dimensional edge contacts,[33] and a metal top gate was added by an Al$_2$O$_3$/Au deposition step.

The Figure 1(a) left inset shows an optical image of the device before the top gate fabrication (located within the dotted rectangle), while the Figure 1(a) right inset shows the layer stacking diagram. Completed devices were measured in a variable-temperature flowing gas $^4$He cryostat. All four-terminal resistance measurements were performed with standard lock-in methods at a base temperature ~1.5 K unless noted. We used the highly-doped silicon substrates as a global bottom gate. In conjunction with the top gate, this device geometry allows independent control of the carrier density [$n = (C_{BG}V_{BG} + C_{TG}V_{TG})/e - n_0$, where $C_{BG}$ and $C_{TG}$ are the bottom and top gate capacitance per area, $V_{BG}$ and $V_{TG}$ are the bottom and top gate voltage, respectively, $e$ is the electron charge and $n_0$ is residual charge due to doping] and the applied displacement field [$D = (C_{BG}V_{BG} - C_{TG}V_{TG})/2\varepsilon_0 - D_0$, where $D_0$ is the residual displacement field].

The longitudinal resistance $R_{xx}$ vs. $V_{BG}$ with $V_{TG} = 0$ is shown in the Figure 1(a) main panel. From this data, we determine the carrier mobility to be ~110,000 cm²V⁻¹s⁻¹ at low temperature, comparable to the high mobility achieved in typical BLG/BN systems. Under the application of a magnetic field, a color plot of $R_{xx}$ vs. $n$ and $B$ shows a Landau fan pattern [Figure 1(b)], consistent with the QHE in bilayer graphene observed previously [see Supporting Information (SI) Figure S1 for more details]. The primary gaps occur at filling factors $\nu = \pm 4m$, where $m$ is an integer. This is accounted for by the bilayer LL spectrum, given by $E_N = \pm \hbar\omega\sqrt{N(N-1)}$, where $\omega = qB/m$ with $q$ the electric charge and $m$ the effective mass, and $N$ is the orbital quantum number index.[34, 35] These LLs are fourfold degenerate due to spin and valley degeneracies, but are split by perturbations such as interactions and Zeeman splitting.[36, 37] Analyzing the QHE data, we extract the values of gate capacitances $C_{BG}$ = 13.5 nF/cm² and $C_{TG}$ = 72.5 nF/cm². For $B$ larger than 6 T, we observe all integer QHE states, indicating the high quality of our sample.

To investigate the layer-selective spin-orbit proximity coupling, we studied the effects of varying $D$. The Figure 1(c) lower panel shows a color plot of $R_{xx}$ vs. $n$ and $D$, taken at $B$ = 10 T. Similar data at $B$ = 6 and 8 T are shown in SI Figure S2. (An artifact likely due to contact resistance and possible formation of resistive $p$-$n$ junctions causes a downward-sloping feature in the plot obscuring the QHE signatures, especially at high $B$. An example of one of these features is marked by a white arrow. Here, we focus on the data outside this region.) The Figure 1(c) upper panel shows $R_{xx}$ plotted vs. $n$ for $D = 0$, yielding a series of $R_{xx}$ minima due to the LLs in the BLG. All the odd filling states are present at $D = 0$ in contrast to BLG without WSe₂.[27, 28] As $D$ varies at fixed $n$, some of the $R_{xx}$ minima vanish and then reemerge, indicating the closing and reopening of inter-LL gaps at LL crossing points. In Figure 1(d), line traces of $R_{xx}$ minima at $\nu =$ 5 and 6 fillings show peaks at particular $D$ values, enabling extraction of the crossing points. Although the $\nu = 7$ curve does not go through a maximum in selected range, the upwards slope suggests a crossing point exists near $D = -0.15$ V/nm. As we find that $R_{xy}$ is impacted less than $R_{xx}$ by the contact artifact (see Figure S3), to ensure accuracy, we also calculate the derivative of $R_{xy}$ with respect to $n$. This produces similar peaks as $D$ varies, enabling the extraction of the same $D$ values for the LL crossings.

The $N = 0, 1$ LLs in an isolated bilayer are degenerate in the single particle picture except for Zeeman splitting, but interactions lift all degeneracies.[24, 29, 36, 38-41] Adding SOC is expected to modify both the inter-LL gaps and crossing points. We zoom in to the $dR_{xy}/d\nu$ vs. $D$ and $n$ data at $B = 8$ T to extract the asymmetric crossing points for the LLs. Figure 2(a) left panel shows this color plot, in which the crossing points are readily visible, such as that marked by the white arrow. From line traces along fixed $n$ at particular filling factors (for example, $\nu = \pm 3$ shown in the right panel), we find and plot the $D$ values at the crossings in Figure 2(b). For most filling factors, the crossing points have the same $D$ magnitude but with their sign depending on the sign of $\nu$. The exception is $\nu = \pm 2$, in which the $\nu = 2$ state crossing occurs at a smaller $D$ value than $\nu = -2$.

To take initial steps towards understanding this behavior, following ref. 20, we assume a phenomenological single-particle model of SOC incorporating both an Ising term, $H_I = \frac{1}{2}\eta\lambda\sigma_z$ and a Rashba term, $H_R = \frac{1}{2}\lambda_R(\eta\sigma_x s_y - \sigma_y s_x)$, where $\lambda$ is an energy scale describing the out-of-plane spin splitting caused by SOC, $\lambda_R$ is an energy scale describing the in-plane Rashba SOC, the $\sigma_{x,y,z}$ are Pauli matrices for spin, $s_{x,y,z}$ are Pauli matrices for the A and B sublattices, and $\eta = +1$ for states near the K point and -1 for the states near the K' point. All bilayer single particle parameters such as nearest neighbor and interlayer hopping are taken from ref. 20. We also add a Zeeman term $H_Z = -½E_Z\sigma_z$,[20] where $E_Z = g\mu_B B$ is the Zeeman energy with $g \approx 2$ the electron g-factor and $\mu_B$ the Bohr magneton, and we account for the interlayer potential $u$ caused by $D$ (LL energies vs. $u$ for $B = 8$ T are plotted in Figure S4). When $D = 0$, all the degeneracies are lifted as a result of SOC coupling in both the model and in the data as shown in the Figure 1(c) upper panel. In contrast, in non-SOC BLG, K-K' degeneracy yields two doubly degenerate LLs in each quartet so that the gaps vanish for $v = \pm 5$ and $\pm 7$. Moreover, under inversion symmetry the crossing points are symmetric in $D$, which is not the case in the SOC device. Thus, the asymmetry in $D$ and fully broken degeneracies are clear evidence of the inversion symmetry breaking caused by the presence of the WSe$_2$.

We now turn to quantitatively estimating $\lambda$ and $\lambda_R$ from the data. The crossing points for $v = \pm 3$ are expected to be unaffected by Coulomb interactions,[29] and are insensitive to the Rashba coupling.[20] The crossing points of $v = \pm 3$ are relatively insensitive to $B$ [Figure 2(b)]. By comparison to the results of the single particle model, we infer that the K point spin splitting has spin down at higher energy than spin up, and vice versa at the K' point. This implies that $\lambda < 0$, which is predicted to occur for certain ranges of graphene-WSe$_2$ twist angles.[30] Such a negative $\lambda$ is also indicated by the crossings for $v = 3$ and 5 having the same sign of $D$ (see SI Figure S4). From these crossing points, we find $\lambda = -2.2$ meV.

For the Rashba coupling, expected to be positive regardless of twist angle,[30] we measure the crossing points' $D$-values for $v = \pm 5, \pm 6,$ and $\pm 7$. Figure 2(c) plots this data for $B = 6$ T and Figure 2(d) plots the corresponding data for $B = 10$ T. Using $\lambda = -2.2$ meV as obtained above, and using $\lambda_R$ as the only free parameter, we minimize the sum of all the squared deviations between the predicted single particle values and the measured data, obtaining $\lambda_R \sim 20$ meV. Figures 2(c) and 2(d) show the resulting model values using the best fit $\lambda_R$ plotted as open triangles. The single particle model with this value of $\lambda_R$ yields a reasonable agreement to the measured values. However, the crossing points are in principle still potentially affected by Coulomb interactions.

To elucidate the role of Coulomb interactions in the behavior of the system, we determine the transport gaps by measuring the temperature dependence of $R_{xx}$ in a number of valleys vs. $D$. An example of this behavior is shown in Figure 3(a) which shows Arrhenius plots for $\nu = 6$ at $D = 0$ vs. $B$. We then fit the $R_{xx}$ minima against $1/T$ with the equation $R = R_0 \exp\left[-\frac{\Delta - \Gamma}{2k_B T}\right]$ to extract the quantity $\Delta - \Gamma$, where $\Delta$ is the gap and $\Gamma$ (full width at half maximum) is the Landau level width. Figure 3(b) shows the extracted $\Delta - \Gamma$ vs. $B$ for $\nu = 5$, 6, and 7.

To make a comparison to the single particle model above, we consider the rate of change of $\Delta - \Gamma$ with $D$ [i.e. $d(\Delta - \Gamma)/dD = d\Delta/dD$ assuming constant $\Gamma$] near a gap closing. This is insensitive to $\Gamma$, as well as Coulomb interactions within a Hartree-Fock picture. Table 1 shows the measured values for a number of crossings. Values on both sides of the crossings are averaged if both are obtained. We then compute the expected $d\Delta/dD$ at each crossing, and using only $\lambda_R$ as a free parameter minimize the sum of the squared differences between all the expected and measured crossing rates with respect to $\lambda_R$. We find a best fit value of $\lambda_R \approx 15$ meV for the data shown. The computed values using this $\lambda_R$ value are tabulated in Table 1 column 4 which show reasonable agreement to the measured values. However, deviations exist which may, for example, stem from measurement uncertainty in the gap closing rates. This indicates that the $\lambda_R$ obtained from this method should be considered only an estimate characteristically ~15 meV, consistent with the range of previously found values, e.g. refs. 9 and 15.

Using this extracted value of $\lambda_R$, we compute the expected behavior of the $\nu = 5$ gap and plot it as the dashed line in Figure 3(b) without free parameters. Fitting a straight line to the $\nu = 5$ gap data in Figure 3(b) yields a slope of $0.11 \pm 0.02$ meV/T, very close to the theoretical slope of $\approx g\mu_B = 0.115$ meV/T, indicating that Coulomb interactions have only a small effect on the gap. The $\nu = 7$ gap is slightly larger, but similar in magnitude. On the other hand, the fitted slope of 0.245 meV/T of the $\nu = 6$ gap is significantly larger than $g\mu_B$, and also increases with $B$. In a single particle picture, the quartet of LLs associated with these gaps consists of states that can be labeled by a K or K' valley index and spin. The K and K' states are split by the SOC into nearly up and down spin states (with slight canting), which are oppositely directed in the two valleys, effectively. The single particle calculation shows that although the Rashba SOC reduces the total splitting, the $\nu = 5$ gap varies linearly with $B$ with a slope $\approx g\mu_B$, as observed. The measured energy is lower than the predicted single particle splitting, which we attribute a LL width ~ 0.4 meV. The results of the single particle model for all three gaps $\nu = 5$, 6 and 7 are plotted over a wider range of $B$ in the Figure 3(b) inset. While this picture produces reasonable results over the measured $B$ range for $\Delta_5$ and $\Delta_7$, it predicts that $\Delta_6$ starts at a splitting set by the SOC, and decreases with $B$, contrary to what is observed.

We therefore consider a model that includes Coulomb interactions. SOC breaks the K-K' symmetry, leading to potentially different exchange energies in the two valleys. Additionally, there could be an exchange coupling between the K and K' states. However, based on the close agreement in the behavior of the $\Delta_5$ and $\Delta_7$ gaps to the single particle picture we assume this is negligible. Thus, we model the Coulomb interaction with two parameters, which are $J_K$ and $J_{K'}$, corresponding to the exchange energy to flip a spin within a LL in the K or K' valleys, with an effective single particle picture shown schematically in Figure 4(a). The similar energy scales measured for $\Delta_5$ and $\Delta_7$ indicate that $J_K \approx J_{K'}$. We obtain $J_{K'}$ by taking the total gap $\Delta_6$ to be $\Delta_{6\text{-sp}}+J_{K'}$, where $\Delta_{6\text{-sp}}$ is the expected single particle gap at $D = 0$, plotted in the inset to Figure 4(b), and rearranging to find $J_{K'} = \Delta_6-\Delta_{6\text{-sp}}$. Because of the unknown Coulomb interactions outside of the measured range and the uncertainty in the LL width, this quantity is plotted in Figure 4(b) as a differential $\Delta J_{K'}$ relative to its value at $B = 6$ T. The behavior is approximately linear in $B$. While it may be initially expected that $J_{K'}$ is proportional to the Coulomb energy scale $e^2/\varepsilon l_B$, where $l_B$ is the magnetic length and $\varepsilon$ the dielectric constant, this result is in agreement with previous experiments that also find an approximately linear in $B$ Coulomb interaction.[31, 32] This may arise from changes to the dielectric constant with $B$.[42] However, more work will be required to understand the origin of this behavior.

In sum, we have measured high mobility bilayer graphene samples in which the spin-orbit interaction is induced by proximity coupling to WSe$_2$. The Ising parameter is measured to be $\lambda \approx -2.2$ meV, while the Rashba parameter is ~10 meV. The SOC breaks all degeneracies of the LLs at $D = 0$, and while the crossing points vs. $D$ are qualitatively in agreement with a single particle model, both the gaps and the crossing points can be affected by Coulomb interactions. We measure the Coulomb energy corrections for $\nu = 6$ and find it is approximately linear in $B$ over the measured range, in accordance with previous results on non-SOC bilayer graphene. The quantum Hall effect is a useful probe of SOC in bilayer graphene. In future work, this approach may be used to test theories in which both the Ising and Rashba SOC are tunable by the twist angle between the graphene and WSe$_2$,[30, 43] potentially affording a unique approach to controlling SOC in materials.

AUTHOR INFORMATION

**Corresponding Author**

*Email: bockrath.31@osu.edu

**Funding Sources**

This research funded by DOE ER 46940-DE-SC0010597. Growth of hexagonal boron nitride crystals was supported by the Elemental Strategy Initiative conducted by the MEXT, Japan, A3 Foresight by JSPS and the CREST(JPMJCR15F3), JST.

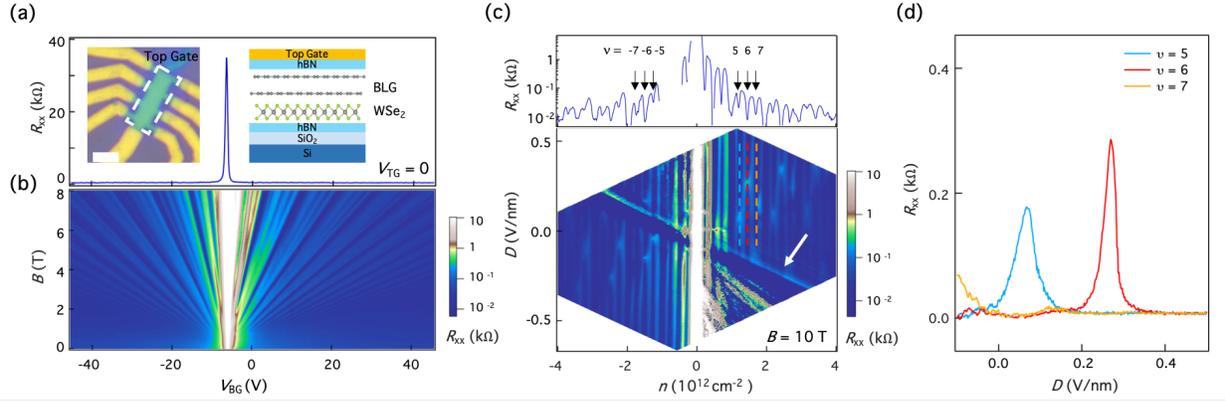

Figure 1. Device structure and SOC modified LL properties. (a) Main panel: $R_{xx}$ vs. bottom gate voltage $V_{BG}$ at $T = 1.5$ K with top gate voltage $V_{TG} = 0$. Left inset: Optical image of a BLG/WSe$_2$ device. The dashed outline indicates the top gated region before metalization. Scale bar is 2 $\mu m$. Right inset: Schematic diagram of the layer stack. (b) $R_{xx}$ vs. $V_{BG}$ and magnetic field $B$ showing a Landau Fan pattern. Quantum Hall states at all integer filling factors of the Landau levels are clearly visible due to the full degeneracy lifting. (c) Top panel: $R_{xx}$ vs. $n$ at $D = 0$. Bottom panel: Color plot of $R_{xx}$ vs. $n$ and $D$ at $B = 10$ T. (d) Vertical line cuts passing Landau Level crossing points along the dashed lines in panel c. Each curve represents a fixed filling factor for $\nu = 5, 6,$ and 7.

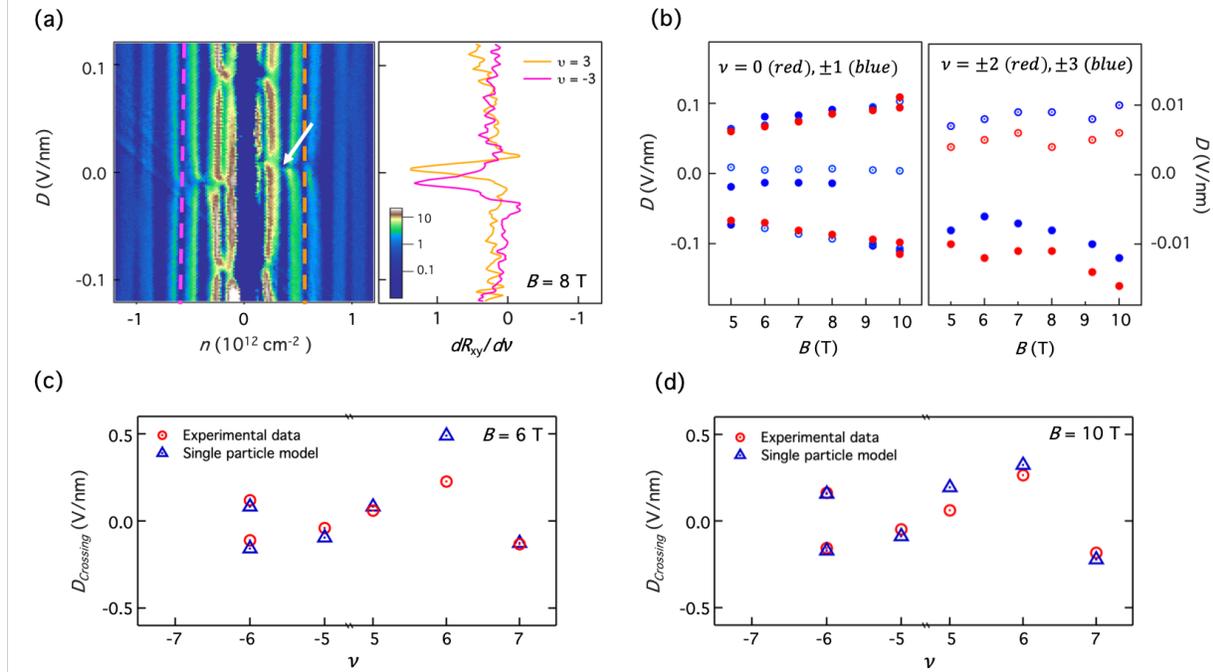

Figure 2. Transitions at zeroth Landau level. (a) Left panel: $dR_{xy}/d\nu$ vs. $n$ and $D$ at $B = 8$ T. Right panel: Vertical line cuts passing LL crossing points along the dashed lines in the left panel at $\nu = \pm 3$. Maxima indicate the crossing points. (b) Measured $D$ values of LL crossing points vs. $B$ for $N = 0$ and 1. Open/filled circles stand for positive/negative filling factors. (c) and (d) show a comparison between measured LL crossing points and our simulated values from the single particle model at $B = 6$ T and 10 T, respectively. [Crossing points at $B = 6$ T are extracted from SI Figure S2(a).]

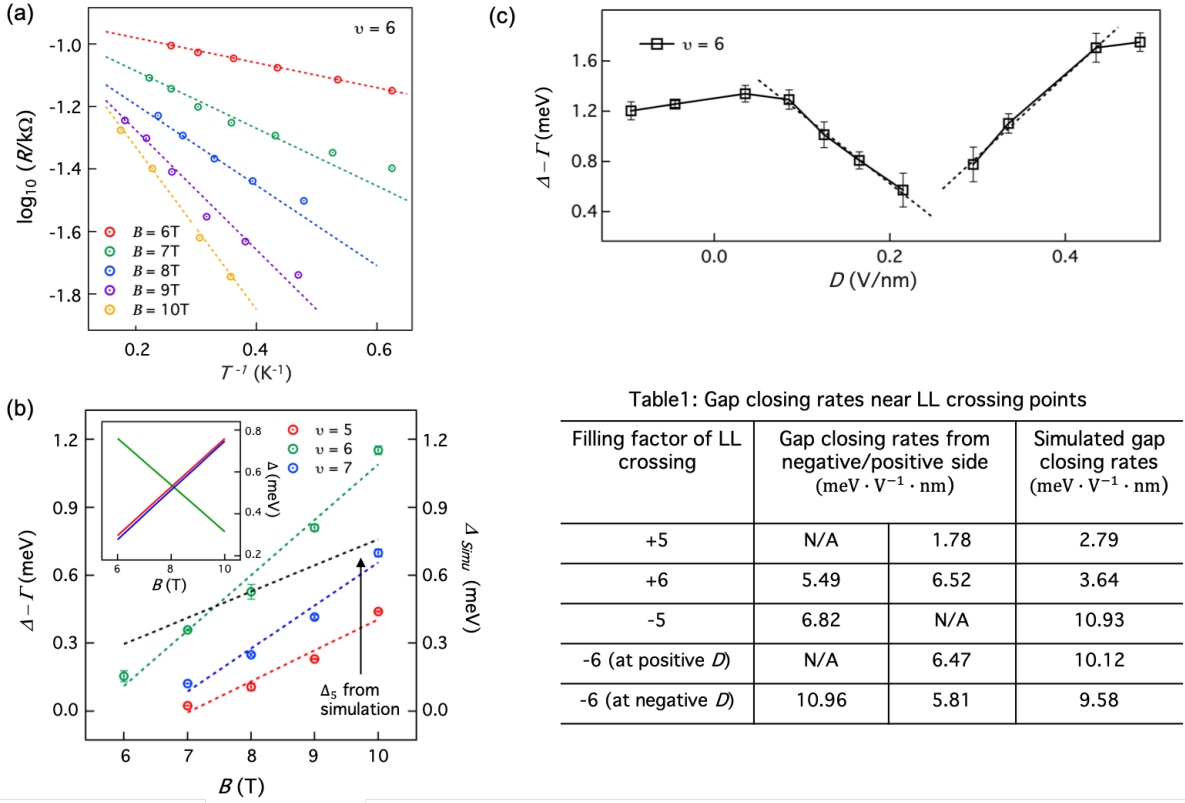

Figure 3. LL energies and broken symmetry gaps in a BLG/WSe$_2$ system. (a) Thermal activation plots at various $B$ values for $\nu = 6$ states at $D = 0$. (b) Measured thermal activation gaps $\Delta - \Gamma$ vs. $B$ at $D = 0$ for $\nu = 5$, 6, and 7, where $\Gamma$ is the LL width (left axis). Dotted lines are linear fits of gap sizes vs. $B$. The black dotted line stands for simulated gap size for $\nu = 5$ without broadening (right axis). Inset: single particle model for all three gaps $\nu = 5$, 6 and 7. Measured gaps $\Delta - \Gamma$ vs. $D$ at $B = 10$ T for $\nu = 6$. The missing data points are at the LL crossing point. By fitting the slopes of the gap changes with $D$ (dashed lines), we extract gap closing rates near LL crossing points and compare them with simulated values, as shown in Table 1. (Gap closing rates for other filling factors in addition to $\nu = 6$ are extracted from data shown in SI Figure S5.)

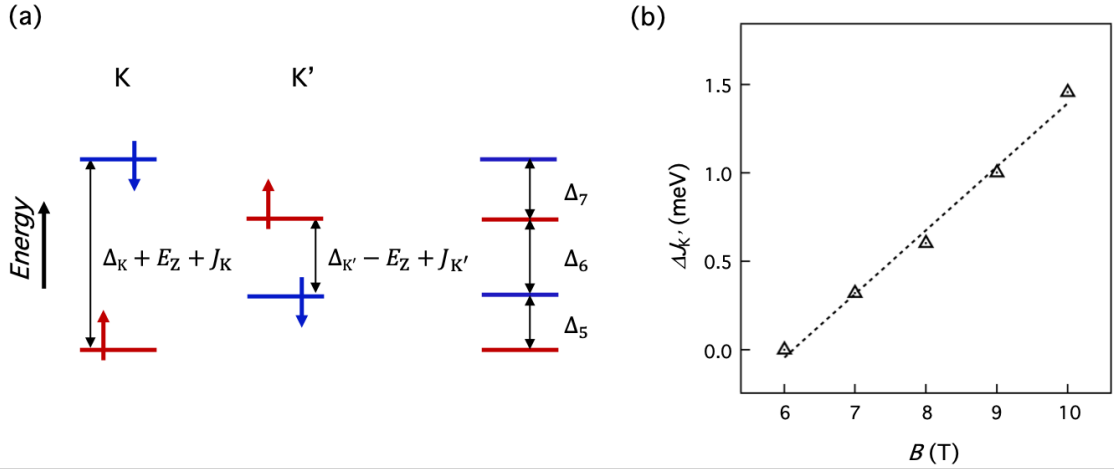

Figure 4. Corrections to the single particle model due to Coulomb interactions. (a) Schematic representation of the broken symmetry gaps for $N = 2$ at $D = 0$. $\Delta_K$ and $\Delta_{K'}$ stands for the proximity induced Ising and Rashba SOC, causing different splitting in K and K' valleys. $J_K$ and $J_{K'}$ are the Coulomb interactions for the electrons with opposite spins in the same valley. (b) Extracted differential $\Delta J_{K'}$ relative to its value at $B = 6$ T, at $D = 0$.

# Supporting Information

# Quantum Hall Effect Measurement of Spin-Orbit Coupling Strengths in Ultraclean Bilayer Graphene/WSe$_2$ Heterostructures

*Dongying Wang[1], Shi Che[1], Guixin Cao[1], Rui Lyu[2], Kenji Watanabe[3], Takashi Taniguchi[3], Chun Ning Lau[1], Marc Bockrath[1]\**

[1] Department of Physics, The Ohio State University, Columbus, OH 43210, USA

[2] Department of Physics and Astronomy, University of California, Riverside, CA 92521, USA

[3] National Institute for Materials Science, Namiki Tsukuba Ibaraki 305-0044 Japan.

*Email: bockrath.31@osu.edu

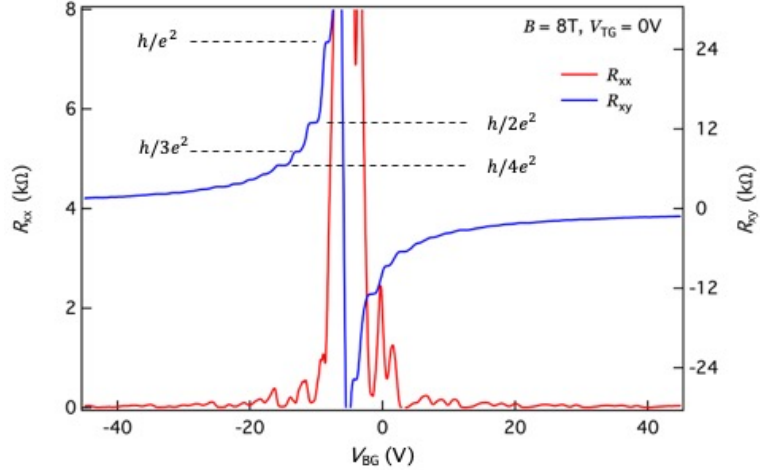

**Figure S1. Integer quantum Hall effect at $B$ = 8 T.** Line cuts from the Landau fan color plot. Integer filling factor features with unit increment, minima in $R_{xx}$ (red) and plateaus in $R_{xy}$ (blue), are visible.

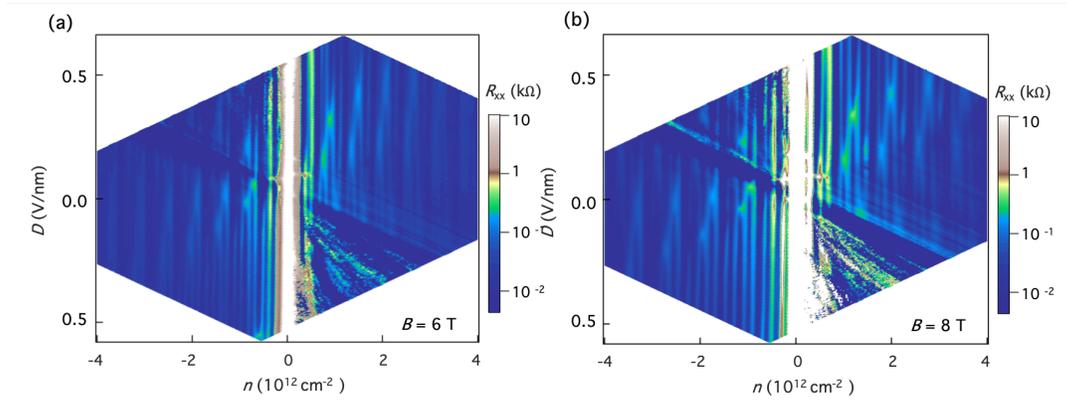

**Figure S2. Color plot of $R_{xx}$ vs. $n$ and $D$ at different magnetic fields.** $R_{xx}$ measurements at additional values of $B$, such as (a) 6T and (b) 8T, in addition to the 10T data we displayed in the main text.

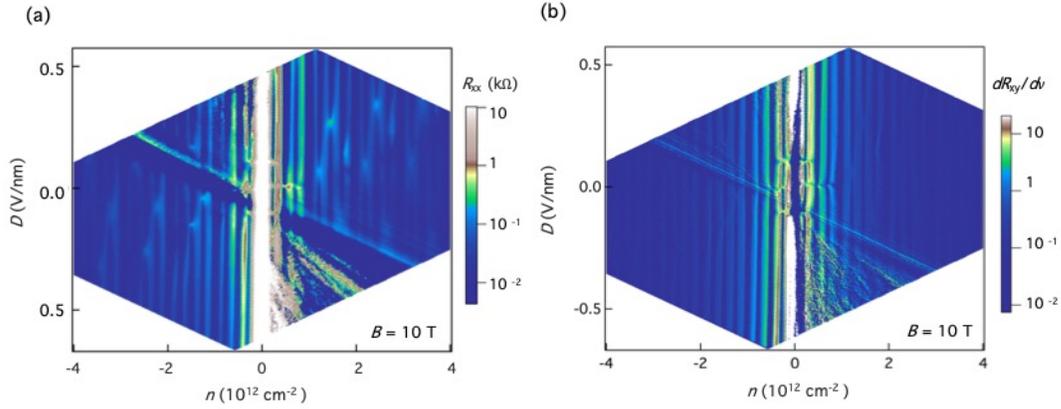

**Figure S3. LL crossing features in color maps of $R_{xx}$ and $dR_{xy}/d\nu$.** (a) Color plot of $R_{xx}$ vs. $n$ and $D$ at $B = 10$ T. (b) Color plot of $dR_{xy}/d\nu$ vs. $n$ and $D$ at $B = 10$ T. From two separate plots, we are able to extract the same information for LL crossings. However, the latter shows the quantum Hall features more clearly, which enable us to extract data more accurately.

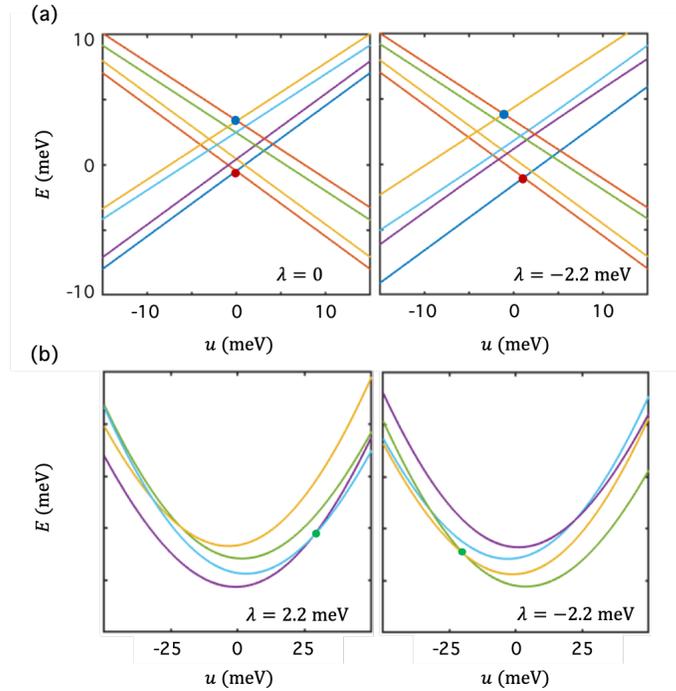

**Figure S4. Energy level diagram in single particle simulations of** (a) Nearly degenerate zero-energy LLs in absence (left) and with (right) Ising type SOC at $B = 8$T. The blue and red dots indicate the crossings for $\nu = +3$ and $-3$, respectively. By comparing the positions of crossing we extract from experimental data [see Figure 3(a)] with simulated results, we find the best fit for value of Ising type SOC is $\lambda = -2.2$ meV. (b) N = 2 LLs with positive (left) and negative (right)

Ising type SOC at $B = 10$T. The green dots indicate the crossings for $\nu = 5$. By comparing the signs of crossing of $\nu = 3$ and 5, we can further confirm that the Ising type SOC is $\lambda = -2.2$ meV.

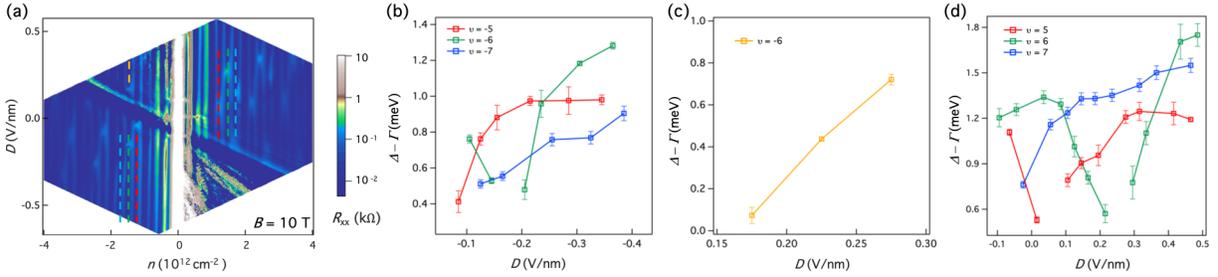

**Figure S5. Displacement field dependence of $N = 2$ QHSs.** (a) Color plot of $R_{xx}$ vs. $n$ and $D$ at $B = 10$ T. (b) and (c) Measured gaps $\Delta - \Gamma$ vs. $D$ at $B = 10$ T for $\nu = -5, -6, -7$ at negative and positive displacement fields. (d) Measured gaps $\Delta - \Gamma$ vs. $D$ at $B = 10$ T for $\nu = 5, 6, 7$ at negative and positive displacement fields. The missing data points are located at LL crossing points. As discussed in the main text, by fitting the slopes of the gap changes with $D$, we extract gap closing rates near LL crossing points and compare them with simulated values.